\documentclass[runningheads]{llncs}
\usepackage{graphicx}

\begin{document}
\title{It's Not Just GitHub: Identifying Data and Software Sources Included in Publications\thanks{Supported by the Alfred P. Sloan Foundation}}
\titlerunning{It's Not Just GitHub}
%
\author{Emily Escamilla\inst{1}\orcidID{0000-0003-3845-7842} \and
Lamia Salsabil\inst{1}\orcidID{0000-0002-6162-2896} \and
Martin Klein\inst{2}\orcidID{0000-0003-0130-2097} \and
Jian Wu\inst{1}\orcidID{0000-0003-0173-4463} \and
Michele C. Weigle\inst{1}\orcidID{0000-0002-2787-7166} \and
Michael L. Nelson\inst{1}\orcidID{0000-0003-3749-8116}}
\authorrunning{E. Escamilla et al.}
%
\institute{Old Dominion University, Norfolk, VA, USA \\
\email{\{evogt001, lsals002\}@odu.edu, \{jwu, mweigle, mln\}@cs.odu.edu} \and
Los Alamos National Laboratory, Los Alamos, NM, USA \\
\email{mklein@lanl.gov}
}
\maketitle              
\begin{abstract}
Paper publications are no longer the only form of research product. Due to recent initiatives by publication venues and funding institutions, open access datasets and software products are increasingly considered research products and URIs to these products are growing more prevalent in scholarly publications. However, as with all URIs, resources found on the live Web are not permanent. Archivists and institutions including Software Heritage, Internet Archive, and Zenodo are working to preserve data and software products as valuable parts of reproducibility, a cornerstone of scientific research. While some hosting platforms are well-known and can be identified with regular expressions, there are a vast number of smaller, more niche hosting platforms utilized by researchers to host their data and software. If it is not feasible to manually identify all hosting platforms used by researchers, how can we identify URIs to open-access data and software (OADS) to aid in their preservation? We used a hybrid classifier to classify URIs as OADS URIs and non-OADS URIs. We found that URIs to Git hosting platforms (GHPs) including GitHub, GitLab, SourceForge, and Bitbucket accounted for 33\% of OADS URIs. Non-GHP OADS URIs are distributed across almost 50,000 unique hostnames. We determined that using a hybrid classifier allows for the identification of OADS URIs in less common hosting platforms which can benefit discoverability for preserving datasets and software products as research products for reproducibility.

\end{abstract}

\section{Introduction}
\begin{figure}
    \centering
    \includegraphics[width=\linewidth]{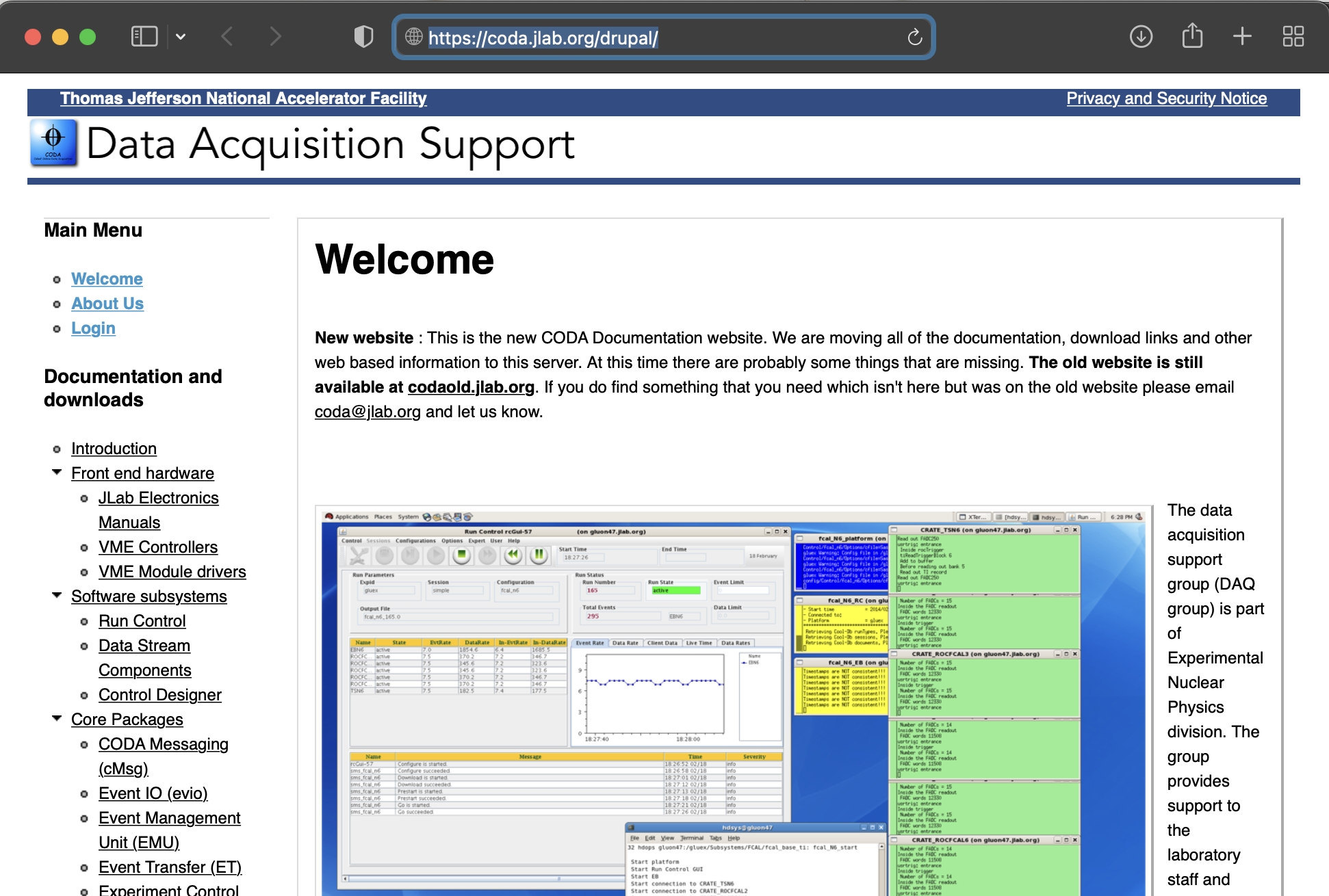}
    \caption{Landing page for https://coda.jlab.org/drupal/}
    \label{fig:jlab}
\end{figure}

The definition of a research product has broadened to include datasets and software products, away from the idea that a paper publication is the only form of research product. In many cases in scientific research, access to the original data and software is the lynch pin of reproducibilty, the ability for other researchers to reproduce or replicate the results of a study. Reproducibility allows for verification of published results as well as further advancement built on the previous methodology. Additionally, publication venues and funding institutions encourage, and in some cases mandate, the sharing of related research objects such as datasets and software. The data and software products produced by researchers, like all Web resources, are subject to content drift \cite{jones-plos2016} and link rot \cite{klein-plos2014}. As such, the notion of access to and reproducibility of research objects is in jeopardy. Therefore, Web archiving efforts are needed to discover, capture, and preserve these research products. However, archivists and archival institutions must be able to find the data and software products created by the researcher in order to preserve them. Some researchers attempt to make their code available for the long-term by self-archiving: depositing their own materials into a repository or archive. However, of academics who write source code, only 47.2\% self-archive that code \cite{iasge_hicss}. Additionally, there is no single platform for researchers to deposit or host their research products. Figure \ref{fig:jlab} shows the Web page where Jefferson Lab, a U.S. Department of Energy Office of Science National Laboratory, hosts the documentation for their CEBAF Online Data Acquisition (CODA) framework. A URI to this Web page was included in over 40 articles in the arXiv\footnote{\url{https://arxiv.org}} corpus, a dataset of 1.58 million STEM pre-print articles used in this study. Both the number of times the URI was included in scholarly publications and the importance of the research institution as a US national laboratory reflect that the size of the platform is incongruent with the importance of its contents. Like the Jefferson Lab, various disciplines and institutions have their own niche hosting platforms, but these may be missed by larger Web preservation efforts like Web archives or Software Heritage due to the relatively small scale of the hosting platforms. 

Software Heritage\footnote{\url{https://softwareheritage.org/}}, a non-profit repository, is solely focused on the preservation of software on the Web with the goal ``to collect, preserve, and share all software that is publicly available in source code form''. They use a combination of crawling and user-submitted requests to discover software products to preserve. Smaller, more niche platforms may be less likely to be discovered resulting in their holdings being less likely to be preserved by Software Heritage without researchers depositing their code. Unlike Software Heritage, Web archives are not solely focused on the preservation of software. Web archives, like Internet Archive, work to archive the Web at large and allow users to experience what a given URI would have looked like at a given point in time. Web archives capture live Web pages, known as URI-Rs, and create a archived copies of the Web page, known as URI-Ms or mementos, with an associated Memento-datetime, the date and time that the URI-R was archived. In their attempt to discover and archive the Web, Web archives may capture Web-based data and software hosting platforms, but this form of incidental archiving is not sufficient for ensuring the preservation of data and software for reproducibility.

There are a number of popular code and data repositories including Zenodo\footnote{\url{https://zenodo.org}}, GitHub\footnote{\url{https://github.com}}, Figshare\footnote{\url{https://figshare.com}}, and Dryad\footnote{\url{https://datadryad.org}}. References to these repositories may be easy to identify by their well-known URIs. We assume that there are other URIs that also point to software and data repositories, so our study is focused on discovering URIs to open-access data and software (OADS) products in order to identify some of the other platforms that scholars are referencing as OADS URIs. We used a classifier system to identify URIs to open access data and software products in scholarly publications. We found that GitHub, GitLab, SourceForge, and Bitbucket account for 33\% of all OADS URIs. We also found that the remaining OADS URIs are distributed across almost 50,000 different Web pages. URIs to the European Council for Nucleur Research (CERN) were the most common with 4,953 URIs, but URIs to CERN still only accounted for 1.92\% of all non-GHP OADS URIs. 

\section{Related Work}
Previous studies have investigated the links between software repositories and scholarly publications. However, most studies have focused their investigations on links to scholarly publications from software repositories. Wattanakriengkrai et al. \cite{wattanakriengkrai_github_2022} studied the extent to which scholarly papers are cited in public GitHub repositories to gain key insights into the landscape of scholarly source code production. A study by Färber \cite{farber-jcdl2020} investigated the characteristics of GitHub repositories included in the Microsoft Academic Graph, which maps code repositories to the research papers they are mentioned in. In previous work \cite{escamilla-tpdl2022}, we looked more broadly at the prevalence of GitHub, GitLab, SourceForge, and Bitbucket in scholarly articles. Like the study by Klein et al. \cite{klein-plos2014} in 2014, they observed a steady rise in the average number of URIs in scholarly publications since 2007 with an average of 5.06 URIs per publication for articles published in 2021. Escamilla et al. also observed an increase in the prevalence of URIs to GitHub, GitLab, SourceForge, and Bitbucket over time with one in five articles including a URI to GitHub alone in 2021. This study will investigate the prevalence of URIs to data and software products and identify the other Web-based repositories and hosting services that scholars are using and citing in scholarly work. 

Scholarly data and software products hosted on the Web are subject to the same ephemeral nature of the Web that plagues URIs to the Web at large. Resources that were once available at a URI may not always be available. This phenomenon is called reference rot. Reference rot is a general term that includes both the effects of link rot and content drift \cite{vandesompel-icm2014}. When a resource identified by a URI is missing, the URI has experienced link rot. When the information identified by a URI changes over time and no longer represents what the author originally intended, the URI has experienced content drift. In a 2014 study, Klein et al. \cite{klein-plos2014} studied the prevalence of reference rot in a dataset of 3.5 million scholarly articles published from 1997 to 2012 from three corpora: arXiv, Elsevier, and PubMed Central. They discovered that one in five articles is impacted by reference rot. For articles that contain at least one URI to the Web at large, seven out of ten articles are impacted by reference rot. Jones et al. \cite{jones-plos2016} took the study a step farther and investigated the prevalence of content drift within the dataset used by Klein et al. They found that 75\% of all URIs referenced in the scholarly articles have been impacted by content drift and no longer reflect what the author originally intended. 

\section{Methodology}
For our study, we analyzed the arXiv corpus. arXiv is a pre-print service for STEM disciplines and its corpus contains over 2 million submissions \cite{fromme-cornell2022}. In April 2007, the arXiv identifier scheme changed to accommodate a larger number of submissions and to address other categorization issues.\footnote{\url{https://arxiv.org/help/arxiv\_identifier}} We decided that beginning our corpus in April 2007 would give us a nearly 15-year time period to study and be sufficient for our analysis. arXiv allows authors to submit multiple versions of their article. In the case of multiple versions, we considered the latest version of each submission to be most representative of what the author intended, so we used only the latest version for our analysis which resulted in a corpus of 1.58 million pre-prints. 

To determine whether a URI links to an open access dataset or software (OADS) resource, we used a hybrid classifier proposed in our previous work \cite{salsabil2022study}. The machine learning classifier works to classify URIs in scholarly publications as OADS or non-OADS. For the study, an OADS URIs was simply defined as a URI linking to open access dataset and/or software. To be classified as OADS, the URI must be open access and a dataset or software product. The classifier transforms each article into a text file using the PDFMiner\footnote{\url{https://pypi.org/project/pdfminer/}} Python library. By employing a regular expression, it scans the text to identify and extract sentences that contain URIs. Given the extracted sentences, the hybrid classifier combines two approaches: a heuristic classifier and a learning-based classifier. The heuristic classifier removes URIs that fall into two categories: those belonging to 54 major publishers such as Springer, Wiley, and Sagepub, and those that end with ``.pdf''. This is because publisher URIs are typically not associated with datasets or software repositories, and .pdf files are typically not datasets or software. The learning-based classifier was trained on a dataset of labeled sentences that contain URIs. The labeled samples were classified as either open access datasets/software (OADS), or not (non-OADS) as shown in Table \ref{tab:classifier}. The OADS cases in the training set were verified to be both open access and data and/or software.The learning-based classifier uses this information to learn how to classify new URIs. In our previous study, we found that the hybrid classifier is more accurate than either the heuristic classifier or the learning-based classifier alone. This is because the heuristic classifier eliminates many non-relevant URIs, and the learning-based classifier is able to accurately classify the remaining URIs.

\begin{table}
    \centering
    \begin{tabular}{|l|c|}
    \hline
    Sentences containing the URI & Category \\ \hline
    The dataset is available at http://ibm.biz/multishapeinsertion. & OADS \\ \hline
    \begin{tabular}[c]{@{}l@{}}Code and materials for reproducing the experiment as well \\ as all data and analysis scripts are open and available at \\ https://github.com/hawkrobe/pragmatics\_of\_perspective\_taking.\end{tabular} & OADS \\ \hline
    \begin{tabular}[c]{@{}l@{}}The codebase that we adapted was developed by Laurent Haan \\ (https://github.com/haan/Lightbot )\end{tabular} & OADS \\ \hline
    \begin{tabular}[c]{@{}l@{}}This article is available from: \\ http://www.nature.com/articles/srep01037.\end{tabular} & Non-OADS \\ \hline
    \begin{tabular}[c]{@{}l@{}}All these scenes can be seen in our video at \\ https://youtu.be/RcWHXL2vJPc.\end{tabular} & Non-OADS \\ \hline
    \begin{tabular}[c]{@{}l@{}}Their contributions are individually acknowledged at\\ http://www.galaxyzoo.org/volunteers.\end{tabular} & Non-OADS \\ \hline
    \end{tabular}
    \caption{Sentences containing OADS and non-OADS URLs.}
    \label{tab:classifier}
\end{table}

After all of the URIs have been classified, we filtered out URIs that were out of scope for this study. In following the methods used by Klein et al. \cite{klein-plos2014} to identify URIs to the ``Web at large'', we filtered out URIs with a scheme other than HTTP or HTTPS, localhost, and private/protected IP ranges. Because we wanted to focus on data and software repositories, we filtered out URIs that would likely point to publications such as URIs to arXiv, Elsevier RefHub,\footnote{\url{https://refhub.elsevier.com}} CrossRef Crossmark \cite{hendricks-crossref-2020}, and some HTTP DOIs. DOIs resolve to artifacts, most commonly papers. Because we were working to identify URIs to data and software, we chose to include DOIs to Zenodo, Dryad, figshare, and Open Science Framework (OSF), as they are known to resolve to data and software artifacts, while removing all other DOIs. Links to Elsevier RefHub and CrossRef Crossmark function similarly to DOIs and are often added by the publisher. We decided to exclude DOI and DOI-like references following Klein et al.'s assumption that, for the most part, such artifacts are in-scope for existing archiving and preservation efforts such as LOCKSS \cite{reich-dlib2001}, CLOCKSS \cite{reich-serials2008}, and Portico \cite{fenton-serials2006}.

We used the regular expressions introduced in our previous study \cite{escamilla-tpdl2022} to identify URIs to GitHub, GitLab, SourceForge and Bitbucket from the extracted URIs. Collectively, we will refer to URIs to one of these four Git hosting platforms (GHPs) as GHP URIs. OADS URIs that are not URIs to one of these four GHPs will be referred to as non-GHP OADS URIs. 

\section{Results}
\begin{figure}
    \centering
    \includegraphics[width=\linewidth]{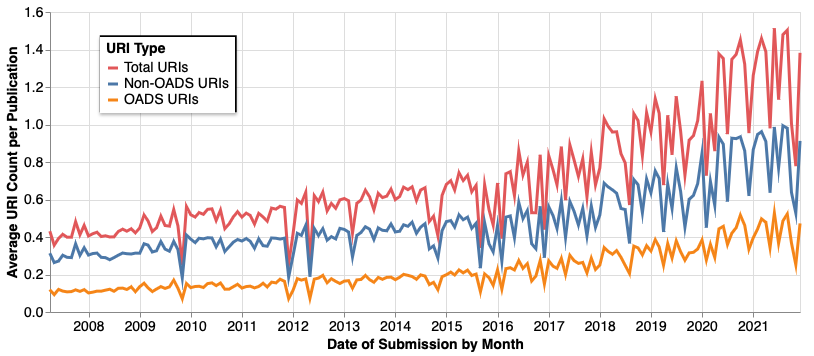}
    \caption{Average number of URIs per arXiv pre-print by publication date. The blue line represents the number of URIs our machine learning model classified as non-OADS as an average per publication (y-axis) per publication month (x-axis). The orange line represent the number of URIs our machine learning model classified as OADS as an average per publication. The red line represents the total number of URIs we extracted from the publications as an average per publication.}
    \label{fig:arxiv_average}
\end{figure}

With the extracted and classified URIs, we looked at the overall distribution of URIs and the distributions of URIs classified as OADS and non-OADS. In Figure \ref{fig:arxiv_average}, we looked at the average number of OADS, non-OADS, and total URIs per publication. The average number of URIs, OADS URIs, and non-OADS URIs per publication rose steadily from 2007 to 2021. In 2007, there were an average of 0.416 URIs per publication with 0.111 OADS URIs per publication and 0.306 non-OADS URIs per publication. Those averages nearly tripled across all three categories by 2021. In 2021, there were an average of 1.273 URIs per publication with 0.433 OADS URIs per publication and 0.841 non-OADS URIs per publication. This shows that authors have been increasingly including URIs, both OADS and non-OADS URIs, in their publications. With a growing number of included URIs comes a growing need to archive the resources that these authors are including in their research with the understanding that authors included the URIs because they were important to their study or were a result of their research.

\begin{figure}
    \centering
    \includegraphics[width=\linewidth]{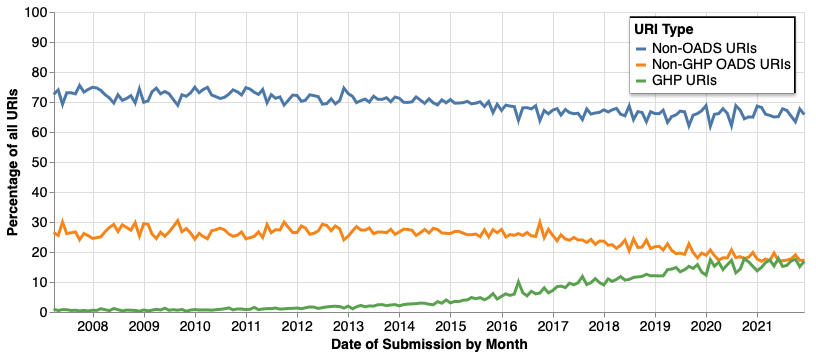}
    \caption{Percentage of GHP URIs, non-GHP OADS URIs, and non-OADS URIs by publication date. The blue line represents the percent of URIs our machine learning model classified as non-OADS (y-axis) per publication month (x-axis). The orange line represents the percent of URIs our machine learning model classified as OADS, excluding GHP URIs. The green line represents the percent of URIs that were GHP URIs.}
    \label{fig:arxiv_percent}
\end{figure}

With an understanding of the general trends of URI usage, we next looked at the distribution of OADS and non-OADS URIs. We also separated the GHP URIs from the other OADS URIs to gain an understanding of the prevalence of GHP URIs over time. We chose GitHub, GitLab, Bitbucket, and SourceForge as popular GHPs to represent GHP URIs. As shown in Figure \ref{fig:arxiv_percent}, we found that both the prevalence and the distribution of the URIs changed across the time period. The percentage of non-OADS URIs has slightly declined meaning that authors are including a higher proportion of OADS URIs to non-OADS URIs in recent years. The percentage of GHP URIs has significantly increased from less than 1\% in 2007 to around 15\% of all URIs in 2021. Despite the overall increase in the prevalence of OADS URIs seen in Figure \ref{fig:arxiv_average}, there has been a decrease in the percentage of non-GHP OADS URIs as shown in Figure \ref{fig:arxiv_percent}. This means that the growth in the prevalence of OADS URIs has largely been due to an increase in the inclusion of GHP URIs within publications. 

\begin{figure}
    \centering
    \includegraphics[width=\linewidth]{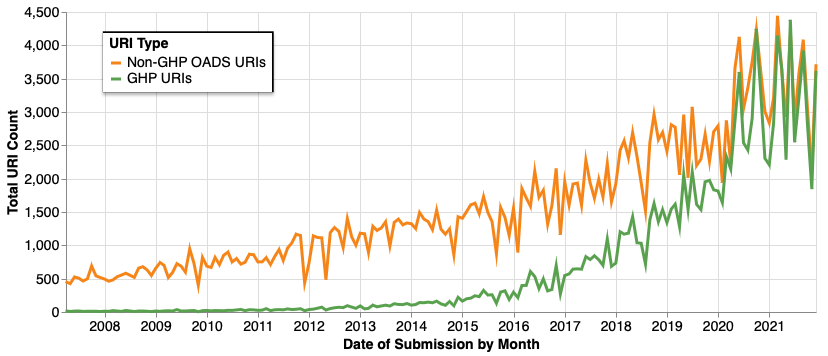}
    \caption{Total number of GHP URIs and non-GHP OADS URIs by publication date}
    \label{fig:arxiv_total}
\end{figure}

The increase of the prevalence of GHP URIs is also reflected when we look at the total number of GHP and OADS URIs over time in Figure \ref{fig:arxiv_total}. From 2007 to 2015, there were a 500 to 1000 more non-GHP URIs than GHP URIs. In 2015, the number of GHP URIs started to steadily increase. In 2020 and 2021, for every GHP URI, there is a non-GHP OADS URI. This shows that utility of using a classifier to identify OADS URIs, especially in older publications from 2007 to 2015. We also see that, while GitHub is an independently popular GHP, we must look beyond GitHub to identify and discover the full breadth of OADS resources being referenced and produced by researchers even in recent year. 

\begin{figure}
    \centering
    \includegraphics[width=\linewidth]{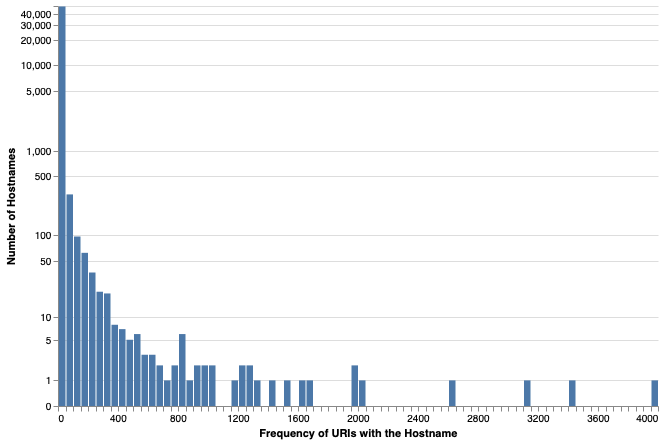}
    \caption{A histogram showing the frequency of the hostname in non-GHP OADS URIs (x-axis) and the number of hostnames that shared that frequency (y-axis).}
    \label{fig:frequency}
\end{figure}

After seeing the trends over time, we wanted to identify the most common non-GHP OADS URIs. We chose to compare URI hostnames and the frequency of those hostnames to determine the most common OADS websites outside of GHPs. In total, we found 258,288 non-GHP OADS URIs included in arXiv publications and almost 50,000 unique hostnames\footnote{The full dataset is available at \url{https://github.com/elescamilla/Extract-URLs/blob/main/classifier\_results/count\_oads\_non\_ghp\_hostnames.csv}} within those URIs. Figure \ref{fig:frequency} shows that 49,392 hostnames are included in between 0 and 50 non-GHP OADS URIs. We found that 63\% of non-GHP OADS URIs are the only URIs to that hostname and only 10\% of URIs reference a hostname that is referenced more than five times. Even with a large number of hostnames referenced a few number of times, there are 19 hostnames that were referenced over 1000 times. Table \ref{tab:hostnames} shows the the top fifteen most common hostnames of non-GHP OADS URIs. However, it is worth noting that the most popular hostname, cds.cern.ch, only accounts for 1.92\% of all non-GHP OADS URIs. Therefore, there are a large number of platforms used by scholars to host data and software which increases the difficulty of archiving data and software products for reproducibility. The diversity of the platforms used and referenced by scholars makes it difficult to manually identify OADS URIs and lends itself to automation like we used with the machine learning classifier model. 

\begin{table}
    \centering
    \begin{tabular}{|l|c|}
    \hline
    Hostname & Frequency \\
    \hline
    cds.cern.ch & 4,953 \\
    www.sciencedirect.com & 3,119 \\
    archive.ics.uci.edu & 2,632 \\
    adsabs.harvard.edu & 2,031 \\
    www.ncbi.nlm.nih.gov & 1,998 \\
    www.cosmos.esa.int & 1,996 \\
    physics.nist.gov & 1,651 \\
    fermi.gsfc.nasa.gov & 1,627 \\
    heasarc.gsfc.nasa.gov & 1,500 \\
    cran.r-project.org & 1,446 \\
    doi.org & 1,337 \\
    www.w3.org & 1,289 \\
    www.nature.org & 1,275 \\
    archive.stsci.edu & 1,243 \\
    en.wikipedia.org & 1,228 \\
    \hline
    \end{tabular}
    \caption{The top 15 most common hostnames for non-GHP OADS URIs and their frequencies.}
    \label{tab:hostnames}
\end{table}

\section{Discussion}

Our analysis found that a significant portion of OADS URIs are not GHP URIs. This shows that, while it is simple to search for OADS URIs by regular expression, regular expressions cannot detect all OADS URIs. Additionally, while the top fifteen most common hostnames are popular platforms for research artifacts, a majority of OADS URIs were to platforms archivists may not know to look for like \url{www.physics.wisc.edu}, \url{www.broadinstitute.org}, \url{fuse.pha.jhu.edu}. Using a classification system like we used in this study, allows us to cast a broader net and detect OADS URIs to lesser known platforms for preservation. While GitHub, GitLab, SourceForge, and Bitbucket accounted for 127,529, or 33\%, of the 385,817 OADS URIs extracted from the arXiv corpus, focusing archival efforts on these and other popular GHPs would miss 67\% of the OADS resources included by researchers. 

\begin{figure}
    \centering
    \includegraphics[width=\linewidth]{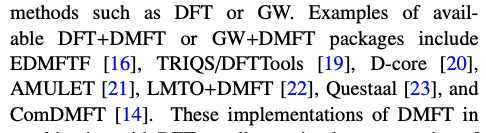}
    \caption[Citation of the ScienceDirect publication.]{Citation of the ScienceDirect publication. The author is listing out available software packages and includes a citation to the ScienceDirect publication and URI.}
    \label{fig:citation}
\end{figure}

\begin{figure}
    \centering
    \includegraphics[width=\linewidth]{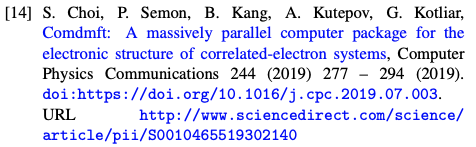}
    \caption[Reference for a ScienceDirect publication]{The reference for a ScienceDirect publication cited in an arXiv publication and classified as an OADS URI despite being a paper publication.}
    \label{fig:reference}
\end{figure}

While the delineation between OADS and non-OADS may seem clear at first glance, it is more nuanced when we look at current citation trends. For example, authors may reference a publication that introduces or discusses a dataset or software product instead of including a direct link to the hosting platform itself. This tendency may be due to the value of publication citations within academia or due to established practices within a discipline or institution, but it results in the possibility of indirect links to OADS via paper publications. For example, ScienceDirect is a digital library of journal articles and book chapters which are non-OADS, but indirect links to OADS could result in ScienceDirect URIs being classified as OADS. Figure \ref{fig:citation} shows a reference to a ScienceDirect publication being cited in an arXiv article in the context of the author listing out available packages for solving DMFT equations. Figure \ref{fig:reference} shows the reference for the ScienceDirect publication. The ScienceDirect publication was classified as an OADS URI by our machine learning classifier model despite it being a paper publication. While the citation itself is to a paper publication, the author is using the citation to indicate a software package discussed in the publication. 

Our machine learning classifier model, despite good performance found in previous studies \cite{salsabil2022study}, fine tuning, and a large training set, was not perfect, as can be expected when extracting and classifying millions of URIs from 1.58 million scholarly articles. It incorrectly classified some GHP URIs as Non-OADS. In some cases, these GHP URIs were located in the footnote or in other locations that lacked the necessary context sentence around the target URI for proper classification. Despite the limitations and inaccuracies, we are remain confident that utilizing machine learning models to classify OADS and non-OADS URIs will allow researchers and archivists to more easily identify less popular or niche platforms for preservation.

\section{Conclusion}

Researchers are increasingly including URIs to the Web at large and also to open access data and software (OADS). However, the multitude of hosting platforms, including institutional or discipline-specific platforms, available to researchers makes it more difficult for archivists to identity these platforms and archive their contents to facilitate long-term reproducibility. We used a machine learning classification system to identify OADS URIs outside of the popular GHP URIs that can be found with regular expresssions. We found that GHP URIs only account for 33\% of all OADS URIs and that non-GHP OADS URIs are dispersed across nearly 50,000 unique hostnames.

\bibliographystyle{splncs04}
\bibliography{paper}

\end{document}